\documentclass[aps,preprint,showpacs,preprintnumbers,floatfix,eqsecnum,prb]{revtex4-1}
\usepackage{amsfonts,amsmath,amssymb}
\usepackage{hyperref}

\usepackage[pctex32]{graphicx}
\usepackage{dcolumn}
\usepackage{bm}
\usepackage{xfrac}

\begin{document}

\newcommand{\Area}{\ensuremath{\mathcal{A}}}
\newcommand{\Dbox}{\ensuremath{\Delta_{\text{Box}} }}
\newcommand{\delh}{\ensuremath{\delta h }}
\newcommand{\epsf}{\epsilon_{\textrm{F}}}
\newcommand{\frth}{\ensuremath{\frac{1}{4}}}
\newcommand{\hef}{$^{4}$\textrm{He }}
\newcommand{\het}{$^{3}$\textrm{He }}
\newcommand{\hetf}{$^{3}$\textrm{He--}$^{4}$\textrm{He }}
\newcommand{\hfz}{\ensuremath{h_{4}^{0}}}
\newcommand{\hz}{\ensuremath{h_{0}}}
\newcommand{\hlf}{\ensuremath{\frac{1}{2}}}
\newcommand{\htm}{\ensuremath{\frac{\hbar^{2}}{2 \mstar}}}
\newcommand{\htp}{\ensuremath{\hbar^{2} 2 \pi}}
\newcommand{\kB}{\ensuremath{k_{\textrm{B}}}}
\newcommand{\kT} {\ensuremath{\kappa_{\Tee}}}
\newcommand{\lthree} {\ensuremath{\ell_{3}}}
\newcommand{\ltsq}{\ensuremath{\lambda_{T}^{2}}}
\newcommand{\mstar}{m^{*}}
\newcommand{\msm}{\ensuremath{m^{*}_{\sigma}}}
\newcommand{\msmm}{\ensuremath{m^{*}_{-\sigma}}}
\newcommand{\msda}{\ensuremath{m^{*}_{\da}}}
\newcommand{\msua}{\ensuremath{m^{*}_{\ua}}}
\newcommand{\muh}{\ensuremath{\mu_{m} \mathcal{H}_{0}}}
\newcommand{\mum}{\ensuremath{\mu_{m}}}
\newcommand{\ntz}{\ensuremath{n_{3}^{0} }}
\newcommand{\nfz}{\ensuremath{n_{4}^{0} }}
\newcommand{\nell}{\ensuremath{n_{\ell}}}
\newcommand{\non}{\ensuremath{n_{\text{onset}}}}
\newcommand{\Numb}{\ensuremath{\mathcal{N}}}
\newcommand{\Tee}{\ensuremath{\mathcal{T}}}
\newcommand{\Pee}{\ensuremath{\mathcal{P}}}
\newcommand{\ua}{\ensuremath{\uparrow}}
\newcommand{\da}{\ensuremath{\downarrow}}

\newcommand{\vf}{\ensuremath{v_{\text{F}}}}
\newcommand{\vfs}{\ensuremath{v_{\text{F}}^{\sigma}}}
\newcommand{\vfsmp}{\frac{v_{\text{F}}^{-\sigma}}{v_{\text{F}}^{\sigma}}}
\newcommand{\vfsp}{\ensuremath{v_{\text{F}}^{\sigma'}}}
\newcommand{\Ns}{\ensuremath{N_{0}^{\sigma}}}
\newcommand{\Nsp}{\ensuremath{N_{0}^{\sigma'}}}
\newcommand{\Ntilde}{\ensuremath{\tilde{N}_{0}}}
\newcommand{\Fzs}{\ensuremath{F^{s}_{0}}}
\newcommand{\Fos}{\ensuremath{F^{s}_{1}}}
\newcommand{\Fza}{\ensuremath{F^{a}_{0}}}
\newcommand{\Foa}{\ensuremath{F^{a}_{1}}}
\newcommand{\gs}{\ensuremath{g_{\sigma}}}
\newcommand{\gu}{\ensuremath{g_{\uparrow}}}
\newcommand{\gd}{\ensuremath{g_{\downarrow}}}
\newcommand{\hu}{\ensuremath{h_{\uparrow}}}
\newcommand{\hd}{\ensuremath{h_{\downarrow}}}
\newcommand{\xis}{\ensuremath{\xi_{\sigma}}}
\newcommand{\xu}{\ensuremath{\xi_{\uparrow}}}
\newcommand{\xd}{\ensuremath{\xi_{\downarrow}}}
\newcommand{\Sigs}{\ensuremath{\Sigma_{\sigma}}}
\newcommand{\Su}{\ensuremath{\Sigma_{\uparrow}}}
\newcommand{\Sd}{\ensuremath{\Sigma_{\downarrow}}}
\newcommand{\nusp}[1]{\ensuremath{\nu^{\sigma}_{#1}}}
\newcommand{\nusm}[1]{\ensuremath{\nu^{-\sigma}_{#1}}}
\newcommand{\mpsm}{\ensuremath{\frac{m^{*}_{\sigma}}{m}}}
\newcommand{\mmsm}{\ensuremath{\frac{m^{*}_{-\sigma}}{m}}}
\newcommand{\nbar}{\ensuremath{\overline{n}}}
\newcommand{\okq}{\ensuremath{\omega_{k q}}}
\newcommand{\kfs}{\ensuremath{k_{\text{F} \sigma}}}
\newcommand{\kfss}{\ensuremath{k^{2}_{\text{F} \sigma}}}
\newcommand{\pol}{\mathcal{P}}
\newcommand{\oomtau}{\ensuremath{\frac{1}{\omega \tau_{\sigma'}}}}
\newcommand{\dis}{\displaystyle}

\newcommand{\kvec}{\mathbf k}
\newcommand{\kvep}{\mathbf k^{\prime}}
\newcommand{\qvec}{\mathbf q}
\newcommand{\pvec}{\mathbf p}
\newcommand{\pvep}{\mathbf p^{\prime}}
\newcommand{\ppq}{\ensuremath{{\mathbf p} + \frac{\mathbf q}{2}}}
\newcommand{\mpq}{\ensuremath{-{\mathbf p} + \frac{\mathbf q}{2}}}
\newcommand{\pppq}{\ensuremath{{\mathbf p^{\prime}} + \frac{\mathbf q}{2}}}
\newcommand{\mppq}{\ensuremath{-{\mathbf p^{\prime}} + \frac{\mathbf q}{2}}}
\newcommand{\mvec}{\mathbf m}
\newcommand{\rvec}{\mathbf r}
\newcommand{\rvep}{{\mathbf r}^{\prime}}

\newcommand{\ssqttwo}{\sin^{2}\left(\frac{\theta_{kk^{\prime}}}{2}\right)}
\newcommand{\tsqttwo}{\tan^{2}\left(\frac{\theta_{kk^{\prime}}}{2}\right)}

\newcommand{\tkkp}{\theta_{k k^{\prime}}}
\newcommand{\ctkkp}{\cos{\tkkp}}
\newcommand{\ctppp}{\cos{\theta_{p p^{\prime}}}}
\newcommand{\cpq}{\cos{\theta_{p q}}}
\newcommand{\stkkp}{\sin{\tkkp}}
\newcommand{\spq}{\sin{\theta_{p q}}}
\newcommand{\ttkkp}{\tan{\tkkp}}
\newcommand{\ctkkptwo}{\ensuremath{ \cos{(\frac{\tkkp}{2}} })}
\newcommand{\stkkptwo}{\ensuremath{ \sin{(\frac{\tkkp}{2}} })}
\newcommand{\ttkkptwo}{\ensuremath{ \tan{(\frac{\tkkp}{2}} })}
\newcommand{\kup}{k_{\uparrow}}
\newcommand{\kdn}{k_{\downarrow}}
\newcommand{\kmkp}{\vert \kvec - \kvep \vert}
\newcommand{\kpkp}{\vert \kvec + \kvep \vert}
\newcommand{\tpq}{\left(\frac{2p}{q}\right)}

\newcommand{\sgn}[1]{\ensuremath{\text{sgn}(#1)}}

%
%

\mathchardef\LL="024C
\def\bra#1{\left\langle#1\right|}
\def\ket#1{\left|#1\right\rangle}
\def\duone{\delta u_1(\rvec)}
\def\duonp{\delta u_1(\rvep)}
\def\duonei#1{\delta u_1(\rvec_#1)}
\def\dutwo#1#2{\delta u_2(\rvec_#1,\rvec_#2)}
\def\ronpi#1{\rho_1(\rvep_#1)}
\def\drone{\delta\rho_1(\rvec)}
\def\dronp{\delta\rho_1(\rvep)}
\def\dronei#1{\delta\rho_1(\rvec_#1)}
\def\drtwo#1#2{\delta\rho_2(\rvec_#1,\rvec_#2)}
\def\dgtwo#1#2{\delta g(\rvec_#1,\rvec_#2)}
\def\difr#1#2{|\rvec_#1-\rvec_#2|}
\def\vph#1#2{V_{\rm p-h}(\rvec_#1,\rvec_#2)}
\def\hm#1{\frac{\hbar^2}{#1m}}
\def\he#1{$^{#1}$He}
\def\half{\frac{1}{2}}
\def\uone{u_1(\rvec)}
\def\uonp{u_1(\rvep)}
\def\uonei#1{u_1(\rvec_#1)}
\def\uonpi#1{u_1(\rvep_#1)}
\def\utwo#1#2{u_2(\rvec_#1,\rvec_#2)}
\def\rone{\rho_4(\rvec)}
\def\ronp{\rho_4(\rvep)}
\def\ronei#1{\rho_4(\rvec_#1)}
\def\rtwo#1#2{\rho_2(\rvec_#1,\rvec_#2)}
\def\gtwo#1#2{g(\rvec_#1,\rvec_#2)}
\def\htwo#1#2{h(\rvec_#1,\rvec_#2)}ARTICLE
\def\comment#1{\bigskip\hrule\smallskip#1\smallskip\hrule\bigskip}
\def\I{{\rm i}}

\title{Competing solutions of Landau's kinetic equation for zero sound and first sound in thin arbitrarily polarized Fermi-liquid films}
\author{David Z. Li}
\email{zhaozhe.li@email.wsu.edu}
\author{R. H. Anderson}
\email{rha@spu.edu}
\author{M. D. Miller}
\email{mdm@wsu.edu}
\author{Ethan Crowell}
\email{ethan.crowell100@email.wsu.edu}
\affiliation{Department of
Physics and Astronomy, Washington State University, Pullman, WA
99164-2814, USA} 

\date{\today}

\begin{abstract}
We examine in detail the method introduced by Sanchez-Castro, Bedell, and Wiegers (SBW) to solve Landau's linearized kinetic equation, and compare it with the well-known standard method introduced by Abrikosov and Khalatnikov (AK).    The SBW approach, hardly known, differs from AK in the way that  moments are taken with respect to the angular functions of the Fourier transformed kinetic equation.    We compare the SBW and AK solutions for zero-sound and first-sound propagation speeds and attenuation both analytically in the zero and full polarization limits, and numerically at arbitrary polarization using Landau parameters appropriate for thin \he3 films.  We find that the lesser known method not only yields results in close agreement with the standard method, but in most cases does so with far less analytic and computational effort. 
\end{abstract}

\maketitle

\section{ \label{sec:Intro} Introduction}

One of the important advances made by Landau when he developed Fermi-liquid theory in the mid-1950's~\cite{Landau56,*Landau57} was the introduction of a kinetic equation that governed the time dependence of the quasiparticle distribution function.  The solution of this kinetic equation yields both the propagation speeds and the attenuation of the collective excitations, among much additional information.  In a recent paper,~\cite{LAM_PRB2013}  we solved Landau's linearized kinetic equation for two-dimensional Fermi liquids with arbitrary polarization, and obtained both analytic expressions and numerically computed results for the propagation speed and attenuation of zero sound and first sound.  The method that we used was the approach pioneered by Khalatnikov and Abrikosov~\cite{Khalatnikov1958,*AK1959}  (AK). This is the classic method that is described in the standard reviews of Landau's Fermi-liquid theory: for example,  Pines and Nozi\`{e}res~\cite{PinesNoz1966} and Baym and Pethick.~\cite{BP1991}   In fact there exists a second approach for solving the kinetic equation that yields results that are similar to but not exactly the same as that of the AK theory but seems to be relatively unknown in the Fermi-liquid literature.  This second approach can be found in the 1989 paper of Sanchez-Castro, Bedell, and Wiegers\cite{CBW1989} (SBW),  and its application to zero sound was discussed briefly in Refs.~\onlinecite{AM2011} and \onlinecite{LAM_PRB2012}.  In this note, we wish to compare in detail the analytic and numerical predictions of the AK and SBW approaches for solving the kinetic equation for the propagation speeds and attenuation of the collective excitations covering both the collisionless and hydrodynamic regimes. To the best of our knowledge such a comparison has not yet been made. In particular, we wish to derive expressions for the propagation speeds and attenuation of both zero sound and first sound as functions of density and polarization.  The attenuation will be determined in the relaxation time approximation for the collision integral. 

In Sec.~\ref{sec:candatt} we shall briefly review the two approaches for solving the kinetic equation.  We shall then use the SBW approach to calculate analytic expressions for zero sound and first sound propagation speeds and attenuation at arbitrary polarization.  These results will be compared and contrasted with our previous analytic results~\cite{LAM_PRB2013} obtained using the AK approach.  Finally, we derive the behavior of the the sound speeds and attenuation in the weak-coupling and strong-coupling limits for both approximations. In Sec.~\ref{sec:he3}, using Landau parameters~\cite{AM2011,LAM_PRB2012,LAM_JLTP2012}  and  quasiparticle lifetimes~\cite{LAM_PRB2013} determined previously, we shall compute sound speeds and attenuation for the system of  thin \he3 films at arbitrary polarization for both approximations, and compare their results.  In Sec.~\ref{sec:Conclusion}, the Conclusion, we shall discuss among other matters the effects on the attenuation and first-sound speeds of using a state-dependent relaxation time approximation for the collision integral. Finally, we shall point out that there are notable advantages of the SBW method especially for systems at non-zero polarization, and also in the zero-sound limit.

\section{\label{sec:candatt}Sound speed and attenuation}

We examine a system of $N = N_{\ua} + N_{\da}$, spin-up and spin-down fermions in a box of area $L^{2}$.  The particles have bare mass $m$, and interact with a typical two-body potential $V(r)$ that is assumed to depend only on the scalar distance between the particles. The energy $E\{n_{\pvec,\sigma}\}$ is a functional of the quasiparticle distribution function.   The particles fill two Fermi seas up to Fermi momenta $\kup$ and  $\kdn$, and we introduce the convention that the spin-down Fermi sea will always be the minority Fermi sea in the case of nonzero polarization. The term \textit{polarization} denotes the magnetization per particle which will be denoted by $\pol$, thus $\pol \equiv M/N = \left( N_{\ua} - N_{\da}\right) / N$.  The system is assumed to be at some finite but low temperature $T$ in the sense that $T << T_{\text{F} \da}$.   The derivation of the dispersion relations, and attenuation of the collective excitations proceeds as in three-dimensions beginning with Landau's linearized kinetic equation:~\cite{BP1991}
\begin{equation} \label{eq:linker}
\frac{\delta  }{\delta t}\delta n_{\pvec,\sigma}(\rvec, t) + \mathbf{v}_{\pvec, \sigma} \cdot \mathbf{\nabla}_{\rvec} \delta n_{\pvec,\sigma} (\rvec, t) - \mathbf{\nabla}_{\pvec}  n_{\pvec,\sigma}^{0} (\rvec, t)  \sum_{\pvec',\sigma'} f_{\pvec \pvec'}^{\sigma \sigma'} \mathbf{\nabla}_{\rvec} \delta n_{\pvec,\sigma} (\rvec, t) = I[ n_{\pvec,\sigma}] \,,
\end{equation}
where $\mathbf{v}_{\pvec, \sigma} \equiv \nabla_{\pvec} \epsilon_{\pvec, \sigma}$ is the Fermi velocity for spin state $\sigma$, $ f_{\pvec \pvec'}^{\sigma \sigma'} = {\delta^{2} E} / {\delta n_{\pvec, \sigma} \delta n_{\pvec^{\prime}, \sigma^{\prime}}}$ are the Landau parameters, and $I[n_{\pvec,\sigma}]$ is the collision integral.   For small oscillations we can assume traveling wave solutions for the excitations: $\delta n_{\pvec,\sigma}(\rvec, t) = \delta n_{\pvec,\sigma} ({\qvec,\omega}) \exp{(i(\qvec \cdot \rvec - \omega t))}$.  The kinetic equation (\ref{eq:linker}) reduces to
\begin{align} \label{eq:linkep}
&\left( q \vfs \cos{\theta} - \omega  \right) \nu_{\sigma}(\theta) + \left( q \cos{\theta} \right) \sum_{\pvec' \sigma'}  f_{\pvec \pvec'}^{\sigma \sigma'}  \delta(\epsilon_{\text{F}}^{\sigma'} - \epsilon_{\pvec' \sigma'}) \vfsp \nu_{\sigma'}(\theta')  \nonumber \\
&= -\frac{1}{i \tau_{\sigma}} \left[  \nu_{\sigma}(\theta) - \langle \nu_{\sigma}(\theta) \rangle - 2 \langle  \nu_{\sigma}(\theta) \cos{\theta} \rangle \cos{\theta} \right] \,,
\end{align}
where $\nu_{\sigma}(\theta)$ is the  Fermi surface distortion for spin state $\sigma$ introduced in the usual way:
\begin{equation}
\delta n_{\pvec,\sigma} (\qvec, \omega) = - \delta(\epsilon_{\text{F}}^{\sigma} - \epsilon_{\pvec \sigma}) \vfs \nu_{\sigma}(\theta) \,,
\end{equation}
and $\theta$ is the angle between $\pvec$ and $\qvec$.  In addition, we have written the collision integral in the relaxation time approximation:
\begin{equation}
\label{eq:CollInt}
I[n_{\pvec, \sigma}] = -\frac{1}{\tau_{\sigma}} \left[ \delta n_{\pvec, \sigma} - \langle \delta n_{\pvec, \sigma} \rangle - 2 \langle \delta n_{\pvec, \sigma} \cos{\theta} \rangle \cos{\theta} \right] \,.
\end{equation}
The second and third terms are added to ensure conservation of particle number, energy, and momentum.\cite{Khalatnikov1958,*AK1959}  The angular brackets are angular averages for two-dimensions: $\langle \dots \rangle \equiv (1/2 \pi) \int_{0}^{2 \pi} \dots d\theta $. Expressions for the quasiparticle lifetimes $\tau_{\sigma}$ were derived in Ref.~\onlinecite{LAM_PRB2013}, and will be briefly discussed below.  There will be additional discussion of the form of Eq.~(\ref{eq:CollInt}) in the Conclusion.

Eq. (\ref{eq:linkep}) can be simplified by introducing Fourier decompositions for the angle-dependent quantities:
\begin{subequations}
\begin{gather}
\nu_{\sigma}(\theta)  = \sum_{\ell = 0}^{\infty} \alpha_{\ell} \nu_{\ell}^{\sigma} \cos{(\ell \theta_{\pvec \qvec})}  \equiv \sum_{\ell = 0}^{\infty} \alpha_{\ell} \nu_{\ell}^{\sigma} T_{\ell}(\cos{\theta_{\pvec \qvec}})  \,, \\
f_{\pvec \pvec'}^{\sigma \sigma'}  = \sum_{\ell = 0}^{\infty} \alpha_{\ell} f_{\ell}^{\sigma \sigma'} \cos{(\ell \theta_{\pvec \pvec'})}  \equiv \sum_{\ell = 0}^{\infty} \alpha_{\ell}  f_{\ell}^{\sigma \sigma'}  T_{\ell}(\cos{(\theta_{\pvec \pvec'})}) \,.
\end{gather}
\end{subequations}
The constants $\alpha_{\ell}$ are defined by
\begin{equation}
\alpha_{\ell} =
\begin{cases}
	1& \text{if $\ell = 0$},\\
	2& \text{if $\ell \ge 1$}.
\end{cases}
\end{equation}
The quantities $T_{\ell}(\cos{\theta}) \equiv \cos{(\ell \theta)}$ are Chebyshev polynomials of the first kind,~\cite{A&S} and were introduced for convenience in Ref.~\onlinecite{LAM_PRB2012}. It was pointed out that integrals over $\theta$ from $0$ to $2 \pi$ can be replaced with integrals over $x \equiv \cos{\theta}$ from $-1$ to $+1$  by simply introducing the weight function $w(x) = 1/\sqrt{1 - x^2}$ and multiplying by a factor of $2$.  This is valid whenever the function involved is real, even in $\theta$, and periodic in $\theta$ with a period of $2 \pi$, which is the case for all  functions needed in this work.  There is no calculational advantage of this second representation. However, when using these variables, expressions in two dimensions become very similar to the familiar expressions in three dimensions by simply substituting Chebyshev polynomials for the usual Legendre polynomials. 
After performing the indicated integrations we find
\begin{align} \label{eq:keq}
(s^{\sigma} - \cos{\theta}) &\sum_{\ell =0}^{\infty} \alpha_{\ell}  \nu_{\ell}^{\sigma} T_{\ell} (\cos{\theta} ) \nonumber \\ 
&- \cos{\theta} \sum_{\sigma'} N^{\sigma'}_{0} \left(\dfrac{\vfsp}{\vfs} \right) \sum_{\ell= 0}^{\infty} \alpha_{\ell} f_{\ell}^{\sigma \sigma'}  \nu_{\ell}^{\sigma'} T_{\ell} (\cos{\theta} ) \nonumber \\
&=  \frac{1}{i \tau_{\sigma} q \vfs } \left(  \nu_{\sigma}(\theta) - \nu_{0}^{\sigma} - 2\nu_{1}^{\sigma} \cos{\theta} \right)  \,,
\end{align}
where  $s^{\sigma} \equiv c/\vfs$, $c \equiv \omega /q $ is the complex valued speed of sound, and $\Ns$ is the \textit{single} spin-state density of states: 
\begin{align} \label{eq:N0s}
\Ns &\equiv   \frac{m^{*}_{\sigma}L^{2}}{2 \pi \hbar^{2}} \,. 
\end{align}
In Eq. (\ref{eq:keq}) the wave vector $q$ is a manifestly complex variable with real and imaginary parts defined by
\begin{equation}
q = q_{1} + i q_{2}  \,,
\end{equation}
where $q_{1}$ and $q_{2}$ are both real.   In the following,  we will obtain for the SBW approximation analytic expressions for the dimensionless speed of sound $s \equiv \omega /( q_{1} v_{\text{F}})$ and the attenuation $\operatorname{Im} (q) \equiv q_{2}$ at arbitrary polarization.  In the limits of zero polarization and full polarization we obtain simple analytic results which can be compared  with those previously obtained with the AK approximation.~\cite{LAM_PRB2013}  The question of whether $s$ is zero sound or first sound will depend on whether $\omega \tau \gg 1$ or $\omega \tau \ll 1$, respectively. Expressions for the quasiparticle lifetimes have been derived in Ref.~\onlinecite{LAM_PRB2013}.

For the AK approach one divides both sides of (\ref{eq:keq}) by $(s^{\sigma} - \cos{\theta})$ and then take moments with respect to the  $T_{\ell} (\cos{ \theta} )$, the angular functions in two dimensions.   We emphasize that despite the presence of the denominator all integrals can be evaluated exactly and analytically.~\cite{LAM_PRB2013}  In practice assuming that the series is truncated after the $\ell =1$ term one obtains a $4 \times 4$ secular equation of substantial complexity  (see Eq. (2.32) in Ref.~\onlinecite{LAM_PRB2013}).  In Ref.~\onlinecite{LAM_PRB2012} it was shown that truncation after the $\ell = 1$ contribution yields accurate sound speeds for a system with \he3  valued Landau parameters.  For the SBW approach one omits the first step of division by $(s^{\sigma} - \cos{\theta})$, and immediately takes moments of (\ref{eq:keq}) with respect to the  $T_{\ell} (\cos{\theta} )$ for $\ell = 0, 1, 2$.   We note that the relevant matrix elements can be found in Ref.~\onlinecite{AM2011}.

There are numerous ways to truncate the AK and SBW sets of linear equations. In a theory capable of yielding both zero sound and spin-zero sound we need to retain deformation coefficients $\nu_{\ell}^{\sigma}$ corresponding to $\ell = 0, 1$.  For the AK set of equations we set $f_{\ell}^{\sigma \sigma'} = 0$ for $\ell \ge 2$.  The deformation parameters $\nu_{\ell}^{\sigma}$ for $\ell \ge 2$ only couple to that Landau parameter with the same value of $\ell$. Thus, with this truncation, no deformation parameter with $\ell \ge 2$ will appear in the truncated AK equations. An example of this can be found in Eqs.~(2.31) in Ref.~\onlinecite{LAM_PRB2013}.

For the SBW equations the procedure differs slightly because the angular integrals couple the deformation parameters differently. In the following we show the first three moments for the SBW equations:
\begin{subequations} \label{eq:kineqtrunc}
\begin{alignat}{2} 
&\ell=0  &s^{\sigma} \nu_{0}^{ \sigma} &=  \nu_{1}^{ \sigma} + \sum_{\sigma'} \left(\dfrac{\vfsp}{\vfs} \right) N^{\sigma'}_{0} f_{1}^{\sigma \sigma'} \nu_{1}^{ \sigma'} \,, \label{eq:u0}\\
&\ell=1   &s^{\sigma} \nu_{1}^{ \sigma} &=  \hlf (\nu_{0}^{ \sigma} + \nu_{2}^{ \sigma})   + \hlf\sum_{\sigma'} \left(\dfrac{\vfsp}{\vfs} \right) N^{\sigma'}_{0} f_{0}^{\sigma \sigma'} \nu_{0}^{ \sigma'}  \,, \label{eq:u1} \\
&\ell=2 \quad  &\left(s^{\sigma} - \frac{ 1}{i \tau_{\sigma} q \vfs } \right) \nu_{2 \sigma} &= \hlf (\nu_{1}^{ \sigma} + \nu_{3}^{ \sigma}) + \hlf \sum_{\sigma'} \left(\dfrac{\vfsp}{\vfs} \right) N^{\sigma'}_{0} f_{1}^{\sigma \sigma'} \nu_{1}^{ \sigma'}  \,. \label{eq:u2}
\end{alignat}
\end{subequations}
In these equations we have applied the same restriction on the Landau parameters that we used for AK: $f_{\ell}^{\sigma \sigma'} = 0$ for $\ell \ge 2$. 

For the SBW set of equations we need to include  the $\ell = 2$ moment.  This is because for the SBW equations particle number conservation and momentum conservation remove the attenuation contribution to the $\ell =0$ and $\ell = 1$ moments.  Thus the $\ell =2$ moment is the lowest order contribution for attenuation to the SBW equations. Further, for SBW, the angular integral for the $\nu_{\ell}^{ \sigma}$ moment couples it to $\nu_{\ell - 1}^{ \sigma}$ and $\nu_{\ell + 1}^{ \sigma}$. Thus the deformation parameter $\nu_{3}^{ \sigma}$ is present in (\ref{eq:u2}).

The value of $\nu_{3}^{ \sigma}$ can be fixed by the following argument. We can proceed in the simplest approximation: zero polarization and $f_{\ell}^{\sigma \sigma'} = 0$ for $\ell \ge 1$, with no loss of generality. In this limit the exact solution to the kinetic equation can be written: 
\begin{equation} \label{eq:nu0theta}
\nu (\theta) \sim \dfrac{\cos{\theta}}{s - \cos{\theta}} \,,
\end{equation}
where we ignore a system dependent constant. In the same limit we can compute $\nu_{3}$. Using Eq.~(3.3a) in Ref.~\onlinecite{AM2011}, and Eqs.(3.23, 3.24, 3.25) in Ref.~\onlinecite{LAM_PRB2012}, we find:
\begin{equation}
\dfrac{\nu_{3}}{s \nu_{0}} = \dfrac{\Omega_{3,0}}{s \Omega_{0, 0}} = z_{0}^{2} = \dfrac{1}{(s + \sqrt{s^{2} - 1})^{2}} \,,
\end{equation}
where the angular integrals $\Omega_{m, n}$ and the parameter $z_{0}$ are defined in Refs.~\onlinecite{AM2011,LAM_PRB2012}.  This ratio equals one in the weak coupling limit ($s \rightarrow 1$), zero in the strong coupling limit ($s \rightarrow \infty$), and it drops monotonically as the interaction strength $f_{0}$ increases. Thus, one can replace $\nu_{3}^{\sigma}$ by $s_{\sigma} \nu_{0}^{\sigma}$ in Eq.~(\ref{eq:u2}), and then Eq.(\ref{eq:u2}) becomes
\begin{equation} \label{eq:nu20}
\nu_{2}^{\sigma} = \nu_{0}^{\sigma} \,.
\end{equation}
This replacement is inaccurate as one approaches the strong-coupling limit.  However, by inspection of Eq.~(\ref{eq:nu0theta}) one can see that in the strong-coupling limit  all of the $\nu_{\ell}$'s vanish except for $\ell =1$, and therefore this inaccuracy is irrelevant.  In Sec.~\ref{ssec:Asym} we shall show that this change to Eq.~(\ref{eq:u2}) gives both the weak and strong-coupling limits correctly.  We note that because  \he3 Landau parameters are not small, \he3 sound speeds are in fact insensitive to the value used for the parameter $\nu_{3}$, and that SBW in particular simply set $\nu_{3} = 0$.

Substituting (\ref{eq:nu20}) into (\ref{eq:u1}) one finds the following eigenvalue equation:
\begin{equation}
\sum_{\sigma''}
\left[ \left( \dfrac{c}{\vfs} \right)^{2} \delta_{\sigma \sigma''} - A_{\sigma \sigma''}\right] \eta_{0 \sigma''} = 0 \,,
\end{equation}
where
\begin{equation} \label{eq:Assp}
A_{\sigma \sigma''} = A_{\sigma \sigma''}' + i A_{\sigma \sigma''}'',
\end{equation}
and we have defined
\begin{subequations} \label{eq:A}
\begin{align} 
A_{\sigma \sigma''}' &\equiv \sum_{\sigma'} \left( \delta_{\sigma \sigma'} + \Nsp f_{1}^{\sigma \sigma'}  \right) \hlf \left[  \left(\dfrac{ 2 + \frac{1}{(\omega \tau_{\sigma'})^{2}}}{ 1 + \frac{1}{(\omega \tau_{\sigma'})^{2}}}  \right) \delta_{\sigma' \sigma''} +  N_{0}^{\sigma''} f_{0}^{\sigma' \sigma''} \right] \left(\dfrac{\vfsp}{\vfs} \right) \,,  \\
A_{\sigma \sigma''}'' &\equiv  - \sum_{\sigma'} \left( \delta_{\sigma \sigma'} + \Nsp f_{1}^{\sigma \sigma'}  \right)   \hlf  \left( \dfrac{ \frac{1}{(\omega \tau_{\sigma'})}}{1 + \frac{1}{(\omega \tau_{\sigma'})^{2}}} \right) \delta_{\sigma' \sigma''}  \left(\dfrac{\vfsp}{\vfs} \right) \,.
\end{align}
\end{subequations}
The eigenfunctions are defined by $ \eta_{0 \sigma} \equiv \left(\dfrac{\vfs}{c} \right) \nu_{0 \sigma} $.  The secular determinant is $2 \times 2$ and the eigenvalues are the roots of a quadratic equation:
\begin{equation} \label{eq:csqpm}
c_{\pm}^{2} = \hlf \left( v_{\text{F} \da}^{2} A_{\da \da} +  v_{\text{F} \ua}^{2} A_{\ua \ua}\right) \pm \hlf \sqrt{\left( v_{\text{F} \da}^{2} A_{\da \da} +  v_{\text{F} \ua}^{2} A_{\ua \ua}\right)^{2} - 4 v_{\text{F} \da}^{2}  v_{\text{F} \ua}^{2} \left( A_{\da \da} A_{\ua \ua}   - A_{\ua \da}  A_{\da \ua}\right)} \,.
\end{equation}
The propagation speed and attenuation can be determined from the real and imaginary parts of $c_{\pm}^{2}$. If we compare this quadratic equation with our previous equivalent AK result Eq. (2.32) in Ref.~\onlinecite{LAM_PRB2013}: it is clear that the numerical solution of  (2.14) is much easier, and yet gives very similar results as will be discussed in detail in the next section.

 In Eq.~(\ref{eq:csqpm}) and below we will have need of the explicit components of $A_{\sigma \sigma''}$.  First the real parts:
\begin{subequations} \label{eq:Areal}
\begin{align} 
A_{\ua \ua}' &=  \left( 1 + N_{0}^{\ua}  f_{1}^{\ua \ua}  \right) \hlf \left[  \left(\dfrac{ 2 + \frac{1}{(\omega \tau_{\ua})^{2}}}{ 1 + \frac{1}{(\omega \tau_{\ua})^{2}}}  \right) +  N_{0}^{\ua} f_{0}^{\ua \ua} \right] + \hlf \left(N_{0}^{\da} f_{1}^{\ua \da}\right) \left(N_{0}^{\ua} f_{0}^{\da \ua} \right) \dfrac{\vf^{\da}}{\vf^{\ua}} \,,  \\
A_{\ua \da}' &=  \hlf \left(N_{0}^{\da}  f_{1}^{\ua \da}  \right) \left[  \left(\dfrac{ 2 + \frac{1}{(\omega \tau_{\da})^{2}}}{ 1 + \frac{1}{(\omega \tau_{\da})^{2}}}  \right) +  N_{0}^{\da} f_{0}^{\da \da} \right] \dfrac{\vf^{\da}}{\vf^{\ua}} + \hlf \left(1 + N_{0}^{\ua} f_{1}^{\ua \ua}\right) N_{0}^{\da} f_{0}^{\ua \da}   \,.
\end{align}
\end{subequations}
For $A_{\da \da}'$ and $A_{\da \ua}'$ reverse the spins in $A_{\ua \ua}'$ and $A_{\ua \da}'$, respectively. 
Next, the imaginary parts:
\begin{subequations} \label{eq:Aimag}
\begin{align}
A_{\ua \ua}'' &=  -\hlf \left( 1 + N_{0}^{\ua}  f_{1}^{\ua \ua}  \right)   \left(\dfrac{ \frac{1}{(\omega \tau_{\ua})}}{ 1 + \frac{1}{(\omega \tau_{\ua})^{2}}}  \right)  \,,  \\
A_{\ua \da}'' &=  -\hlf \left(N_{0}^{\da}  f_{1}^{\ua \da}  \right)  \left(\dfrac{ \frac{1}{(\omega \tau_{\da})}}{ 1 + \frac{1}{(\omega \tau_{\da})^{2}}}  \right)  \dfrac{\vf^{\da}}{\vf^{\ua}}    \,.
\end{align}
\end{subequations}
For $A_{\da \da}''$ and $A_{\da \ua}''$ reverse the spins in $A_{\ua \ua}''$ and $A_{\ua \da}''$, respectively. 

In the small attenuation regime $q_{2}/q{1} \ll 1$ we can substitute (\ref{eq:Assp}) into (\ref{eq:csqpm}) and expand to first-order in the $A_{\sigma \sigma''}''$'s:
\begin{align} \label{eq:csq}
c^{2} = (c^{\prime})^{ 2} &+ i \Biggl[ \hlf \left(  v_{\text{F} \ua}^{2} A_{\ua \ua}'' + v_{\text{F} \da}^{2} A_{\da \da}''  \right) \Biggr. \nonumber \\
 &+ \left\{ \hlf \left( v_{\text{F} \da}^{2} A_{\da \da}' +  v_{\text{F} \ua}^{2} A_{\ua \ua}' \right) \left( v_{\text{F} \da}^{2} A_{\da \da}'' +  v_{\text{F} \ua}^{2} A_{\ua \ua}'' \right) \right. \nonumber \\
 &-  \Biggl. v_{\text{F} \da}^{2}  v_{\text{F} \ua}^{2} \left[ A_{\ua \ua}' A_{\da \da}'' +  A_{\da \da}' A_{\ua \ua}''   - A_{\ua \da}'  A_{\da \ua}'' - A_{\da \ua}' A_{\ua \da}''  \right] \Bigr\} \nonumber \\
 &\times \Biggl. \dfrac{1}{\sqrt{\left( v_{\text{F} \ua}^{2} A_{\ua \ua}' +  v_{\text{F} \da}^{2} A_{\da \da}' \right)^{2} - 4 v_{\text{F} \da}^{2}  v_{\text{F} \ua}^{2} \left( A_{\ua \ua}'  A_{\da \da}'   - A_{\ua \da}'  A_{\da \ua}' \right)}} \Biggr] \,.
\end{align}
In this expression, $ (c^{\prime})^{ 2}$ is the real part of the complex speed squared. That is, it is Eq.~(\ref{eq:csqpm}) with $A_{\sigma \sigma'}$ replaced by $A_{\sigma \sigma'}'$.  We have omitted the spin-zero-sound (minus) root because spin-zero-sound does not propagate in \he3 thin films.~\cite{LAM_PRB2012} It is simple to recover that solution by replacing the plus sign by a minus sign in front of the first curly bracket in Eq.~(\ref{eq:csq}), and also before the radical in (\ref{eq:csqpm}). We have defined $c^{\prime} \equiv \omega/q_{1}$ and in lowest order of $q_{2}/q_{1}$ the attenuation can be found from:
\begin{equation} \label{eq:csqrealimag}
c^{2} = (c^{\prime})^{2} \left( 1 - 2i \frac{q_{2}}{q_{1}}  \right) \,.
\end{equation}
Eqs. (\ref{eq:csq}) and (\ref{eq:csqrealimag}) are an analytic expression for the complex speed of sound in the SBW approximation.  In the following we shall use Eq.(\ref{eq:csq}) for obtaining simple analytic expressions  in terms of the Landau parameters in the zero-sound and first-sound limits.  For the numerical work that is discussed in Sec.~\ref{sec:he3}, we use Eq.~(\ref{eq:csqpm}) directly. 

\subsection{ \label{ssec:zero} Zero sound}

In the zero sound limit, we let $\omega \tau_{\sigma} \rightarrow \infty$. Thus, from Eq.~(\ref{eq:A}) we find:
\begin{equation} \label{eq:A0ssp}
A_{\sigma \sigma''}^{(0)} = A_{\sigma \sigma''}^{(0) \prime} + i A_{\sigma \sigma''}^{(0) \prime \prime} \,,
\end{equation}
where to lowest order in $1/\omega \tau_{\sigma}$:
\begin{align} \label{eq:A0sspRe}
 A_{\sigma \sigma''}^{(0) \prime} &= \sum_{\sigma'} \left( \delta_{\sigma \sigma'} + \Nsp f_{1}^{\sigma \sigma'}  \right) \left[ \delta_{\sigma' \sigma''} + \hlf N_{0}^{\sigma''} f_{0}^{\sigma' \sigma''} \right] \left(\dfrac{\vfsp}{v_{\text{F}}^{\sigma}} \right) \,, \\
\label{eq:A0sspIm}
A_{\sigma \sigma''}^{(0) \prime \prime} &=  -\hlf \sum_{\sigma'} \left( \delta_{\sigma \sigma'} + \Nsp f_{1}^{\sigma \sigma'}  \right)  \left(\dfrac{\vfsp}{\vfs} \right) \frac{1}{\omega \tau_{\sigma'}} \delta_{\sigma' \sigma''} \,.
\end{align}
Substituting Eq.~(\ref{eq:A0sspRe}) into Eq.~(\ref{eq:csqpm}), and substituting Eqs.~(\ref{eq:A0sspRe}) and \ref{eq:A0sspIm}) into Eq.~(\ref{eq:csq}) and comparing with Eq.~(\ref{eq:csqrealimag}), one obtains analytic expressions for the speed and attenuation of zero sound, respectively. Comparing with the analogous expressions previously obtained by the AK method, Eqs.~(2.37)--(2.41) in Ref.~\onlinecite{LAM_PRB2013}, the SBW results have a much simpler mathematical structure. This will be seen more clearly when the two limiting cases, zero and full polarization, are discussed below. Despite the fact that the SBW approximation leads to very different \textit{looking} expressions for the speed and attenuation of zero sound, we shall show in Sec.~\ref{sec:he3} below that SBW yields numerical results that are very close to the AK approximation.

\subsubsection{\label{ssec:z0p0}Zero polarization}

In the zero polarization limit $ v_{\text{F}} = v_{\text{F} \ua} =  v_{\text{F} \da}$, and  Eq.~(\ref{eq:csq}) becomes
\begin{equation} \label{eq:csqp0}
s^{2} = s_{0}^{2} +   i \left( A_{\ua \ua}^{(0) \prime \prime}  +  A_{\ua \da}^{(0) \prime \prime} \right)   \,,
\end{equation}
where the zero-sound dimensionless propagation speed $s_{0} = \omega/(\vf q_{1})$ is given by :
\begin{equation}
s_{0}^{2} =  A_{\ua \ua}^{(0) \prime} + A_{\da \ua}^{(0) \prime} \,.
\end{equation}
Using Eq.~(\ref{eq:csq}) we find
\begin{equation}
-2ic_{0}^{2} \dfrac{q_{2}}{q_{1}} = i \left( A_{\ua \ua}^{(0) \prime \prime} + A_{\ua \da}^{(0) \prime \prime} \right) v_{\text{F}}^{2} = -i \hlf \dfrac{1}{ \omega \tau} \left(  1 +  F_{1}^{s} \right) v_{\text{F}}^{2}\,,
\end{equation} 
where:
\begin{align}
A_{\ua \ua}^{(0) \prime} &=  (1 + \hlf F_{0}^{\ua \ua}) (1 + F_{1}^{\ua \ua}) + \hlf F_{1}^{\ua \da} F_{0}^{\da \ua} \nonumber \,, \\
A_{\ua \da}^{(0) \prime} &=  (1 + \hlf F_{0}^{\da \da}) F_{1}^{\ua \da} + \hlf (1 + F_{1}^{\ua \ua}) F_{0}^{\ua \da} \nonumber \,, \\
A_{\ua \ua}^{(0) \prime \prime} &= -\hlf \left( \dfrac{1}{ \omega \tau}  \right) (1 + F_{1}^{\ua \ua}) \nonumber \,, \\
A_{\ua \da}^{(0) \prime \prime} &= -\hlf \left( \dfrac{1}{ \omega \tau}  \right) F_{1}^{\ua \da} \,.
\end{align}
For convenience we have introduced dimensionless Landau parameters defined by
\begin{equation}
F_{\ell}^{\sigma \sigma'} \equiv \Nsp f_{\ell}^{\sigma \sigma'} \,.
\end{equation}
Note that these are defined with the single spin-state density of states, and they are not symmetric in the indices at non-zero polarization.  The symmetric and antisymmetric Landau parameters used in the zero-polarization limit are defined as usual by
\begin{align} \label{eq:symmF}
2 \Nsp f_{\ell}^{\sigma \sigma'} = {F}_{\ell}^{s} + \sigma  \sigma'  {F}_{\ell}^{a}  \,, 
\end{align}
where for this definition we associate $\sigma (\ua) = +1$, and $\sigma (\da) = -1$. The Landau parameters that appear in Eq.~(\ref{eq:symmF}) are the usual  parameters scaled with the \textit{two} spin-state density of states. Two spin-state Landau parameters are only used in the zero-polarization case.

Thus the  speed and attenuation become:
\begin{subequations} \label{eq:zsp0sbw}
\begin{align} 
s_{0}^{2}   &=    ( 1 + \hlf F_{0}^{s}  ) \left(1 + F_{1}^{s}   \right)  \,, \\
\dfrac{q_{2}}{q_{1}} &= \frth   \dfrac{ \left( 1 +  F_{1}^{s} \right)}{s_{0}^{2}} \dfrac{1}{\omega \tau} =  \dfrac{1}{4 ( 1 + \hlf F_{0}^{s}  )} \dfrac{1}{\omega \tau} \,.
\end{align}
\end{subequations}
The expressions for spin-zero-sound can be obtained by replacing the symmetric Landau parameters with antisymmetric Landau parameters. 

The analogous results for the zero-sound speed and attenuation that one obtains from the AK approach~\cite{LAM_PRB2013} are:
\begin{subequations} \label{eq:zsp0ka}
\begin{align} 
g(s_{0}) &= \frac{1 + \Fos}{(1 + \Fos)\Fzs + 2s_{0}^{2}\Fos} \,, \label{eq:c0P0} \\
\frac{q_{2}}{q_{1}}&= \frac{1}{\omega \tau} \Bigg[1 -  \frac{(1 + \Fos)(1 + g(s_{0})) + 2 s_{0}^{2} (1 - \Fos) g(s_{0})}{\frac{\displaystyle (1 + \Fos)}{\displaystyle g(s_{0})} \frac{\displaystyle (1 + g(s_{0}))}{\displaystyle s_{0}^{2} - 1} - 4 s_{0}^{2} \Fos g(s_{0})}  \Bigg] \,. \label{eq:xiprime}
\end{align}
\end{subequations}
The important function $g(s)$ is defined by:
\begin{equation} \label{eq:g}
g(s) \equiv \dfrac{s}{\sqrt{s^{2} - 1}} - 1 \,.
\end{equation}

\subsubsection{Full Polarization}

In the limit of full polarization, we set $ v_{\text{F}} =  v_{\text{F} \ua} \text{ and }  v_{\text{F} \da} = 0$.  Then Eq.~(\ref{eq:csq}) becomes
\begin{equation} \label{eq:csqp1}
c^{2} = c_{0}^{2} +  i A_{\ua \ua}^{(0) \prime \prime}  v_{\text{F} \ua}^{2}\,,
\end{equation}
where the zero-sound propagation speed is $c_{0}^{2} =  A_{\ua \ua}^{0}  v_{\text{F} \ua}^{2}$.  In terms of Landau parameters the dimensionless zero-sound speed $s_{\ua}$  and attenuation are given by:
\begin{subequations} \label{eq:zsp1sbw}
\begin{align}
s_{0} ^{2} &= (1 + \hlf F_{0}^{\ua \ua}) (1 + F_{1}^{\ua \ua})  \,, \\
\dfrac{q_{2}}{q_{1}} &=  \dfrac{1}{4 (1 + \hlf F_{0}^{\ua \ua})} \left( \dfrac{1}{\omega \tau_{\ua}}  \right) \,.
\end{align}
\end{subequations}
The analogous results using the AK approach yield:
\begin{subequations} \label{eq:zsp1ka}
\begin{align} \label{eq:c0P1}
g(s_{\ua}) &=    \frac{1 + F_{1}^{\ua \ua}  }{F_{0}^{\ua \ua} (1 + F_{1}^{\ua \ua}) + 2 s_{\ua}^{2} F_{1}^{\ua \ua}} \,, \\
\dfrac{q_{2}}{ q_{1}} &= \Big[\frac{c_{\tau \ua}^{0} + F_{1}^{\ua \ua} c_{\tau \ua}^{1} }{ F_{0}^{\ua \ua} \hu +   F_{1}^{\ua \ua} c_{q \ua}^{1}  } \Big] \left( \dfrac{1}{\omega \tau_{\ua}}  \right)  \,.
\end{align}
\end{subequations}
The parameters in the attenuation were introduced for convenience in Ref.~\onlinecite{LAM_PRB2013} and are defined by:
\begin{subequations}
\begin{align}
\hu &\equiv s_{\ua} \Big( \frac{\partial \gu}{\partial s_{\ua}} \Big)_{\xi' = 0} \,, \\
\xu' &\equiv \frac{1}{\omega \tau_{\ua}} -  \frac{q_{2}}{q_{1}} \,, \\
 c_{\tau, \sigma}^{0}  &\equiv 1 + (2 s_{\sigma}^{2} + 1) g(s_{\sigma}) + F_{0}^{\sigma \sigma} h_{\sigma} \,, \\
 c_{\tau, \sigma}^{1}  &\equiv 1 + 2 s_{\sigma}^{2} (g(s_{\sigma})  + h_{\sigma}) + g(s_{\sigma})  + F_{0}^{\sigma \sigma} h_{\sigma} \,, \\
 c_{q, \sigma}^{1}  &\equiv 2 s_{\sigma}^{2} (2 g(s_{\sigma})  + h_{\sigma}) + F_{0}^{\sigma \sigma} h_{\sigma}  \,. 
\end{align}
\end{subequations}

\subsection{First sound}

In the hydrodynamic regime $\omega \tau \ll 1$.  Using the same notation as for zero sound, we have from Eq.~(\ref{eq:A}):
\begin{equation} \label{eq:A1ssp}
A_{\sigma \sigma''}^{(1)} = A_{\sigma \sigma''}^{(1) \prime} + i A_{\sigma \sigma''}^{(1) \prime \prime} \,,
\end{equation}
where to lowest order in $\omega \tau_{\sigma}$:
\begin{subequations} \label{eq:FsApApp}
\begin{align}
 A_{\sigma \sigma''}^{(1) \prime} &= \hlf \sum_{\sigma'} \left( \delta_{\sigma \sigma'} + \Nsp f_{1}^{\sigma \sigma'}  \right)  \left[ \delta_{\sigma' \sigma''} + N_{0}^{\sigma''} f_{0}^{\sigma' \sigma''} \right] \left(\dfrac{\vfsp}{v_{\text{F}}^{\sigma}} \right)  \label{eq:A1real} \,, \\
A_{\sigma \sigma''}^{(1) \prime \prime} &=  -\hlf \sum_{\sigma'} \left( \delta_{\sigma \sigma'} + \Nsp f_{1}^{\sigma \sigma'}  \right)  \left(\dfrac{\vfsp}{\vfs} \right) \left( \omega \tau_{\sigma'} \right) \delta_{\sigma' \sigma''} \,.
\end{align}
\end{subequations}
The general expression for the square of the complex first sound speed is the same as Eq.~(\ref{eq:csq}) with the $0$ superscripts replaced by $1$'s. 

For the AK approach, it was shown in Eq.~(2.42) of Ref.~\onlinecite{LAM_PRB2013} that the first sound speed could be written:
\begin{equation} \label{eq:c1P}
\begin{split}
&\Big[  2  (\dfrac{c_{1}}{v_{\text{F}}^{\ua}})^{2} - (1 + F_{0}^{\ua \ua})  (1 + F_{1}^{\ua \ua}) \Big] \Big[  2  (\dfrac{c_{1}}{v_{\text{F}}^{\da}})^{2} - (1 + F_{0}^{\da \da})  (1 + F_{1}^{\da \da}) \Big]  \\
- &(F_{0}^{\ua \da})^{2} \Big[  (1 + F_{1}^{\ua \ua})  (1 + F_{1}^{\da \da})  -  (F_{1}^{\ua \da})^{2}\Big] - (F_{1}^{\ua \da})^{2}  (1 + F_{0}^{\ua \ua})  (1 + F_{0}^{\da \da})  \\
- &4 F^{\ua \da}_{0} F^{\ua \da}_{1} \dfrac{c_{1}}{v_{\text{F}}^{\ua}} \dfrac{c_{1}}{v_{\text{F}}^{\da}} = 0 \,.
\end{split}
\end{equation}
It was surprising to discover that after some algebra (\ref{eq:c1P}) was found to be identical to Eq.~(\ref{eq:csq}) together with Eq.~(\ref{eq:A1real}).  Thus the SBW and AK methods yield identical results for the first-sound speeds at all polarizations. As will be seen below this equality is not carried over to the attenuation.

\subsubsection{Zero polarization}

At zero polarization $s^{2} = s_{1}^{2} + i (A_{\ua \ua}^{(1)  \prime \prime} + A_{\ua \da}^{(1)  \prime \prime})$. Thus, we find for the dimensionless first-sound propagation speed $s_{1} \equiv \omega / (q_{1} \vf)$, and the attenuation: 
\begin{subequations} \label{eq:fsp0sbw}
\begin{align} 
s_{1}^{2} &=  A_{\ua \ua}^{(1)  \prime} + A_{\ua \da}^{(1)  \prime} =\hlf \left(1 +  F_{0}^{s}\right) \left(1 + F_{1}^{s}   \right) \,, \\
\dfrac{q_{2}}{q_{1}} &=    \dfrac{\omega \tau }{2 ( 1 +  F_{0}^{s} )} \label{eq:q2SBWP0}\,.
\end{align}
\end{subequations}
Using AK theory, we find:
\begin{subequations} \label{eq:fsp0ka}
\begin{align}
\label{eq:c1P0}
s_{1}^{2} &= \hlf (1 + \Fzs) (1 + \Fos)  \,, \\
\label{eq:q2AKP0}
\dfrac{q_{2}}{q_{1}} &= \frac{\omega \tau}{4 (1 + \Fzs)} \,.
\end{align}
\end{subequations}

\subsubsection{Full polarization}

In the limit of full polarization, we set $ v_{\text{F}} =  v_{\text{F} \ua} \text{ and }  v_{\text{F} \da} = 0$.  Thus we obtain $c^{2} = c_{1}^{2} +  i A_{\ua \ua}^{(1) \prime \prime}  v_{\text{F} \ua}^{2}$, and therefore
\begin{subequations} \label{eq:fsp1sbw}
\begin{align} 
s_{1} ^{2} &=  \hlf (1 + F_{0}^{\ua \ua}) (1 + F_{1}^{\ua \ua}) \,, \\
\dfrac{q_{2}}{q_{1}} &=  \hlf \dfrac{\omega \tau_{\ua}}{(1 + F_{0}^{\ua \ua})} \label{eq:q2SBWP1} \,.
\end{align}
\end{subequations}
In the AK approach we found:
\begin{subequations} \label{eq:fsp1ka}
\begin{align} 
s_{1}^{2} &= \frac{1}{2}  \left( 1+ F_{0}^{\ua \ua} \right) \left( 1+ F_{1}^{\ua \ua} \right)  \,, \\
\dfrac{q_{2}}{ q_{1}} &= \frac{ \omega \tau_{\ua}}{4 ( 1+ F_{0}^{\ua \ua} ) } \label{eq:q2AKP1}  \,.
\end{align}
\end{subequations}
Thus, in the zero and full polarization limits the SBW and AK attenuation differ by a factor of two. 

We can obtain additional information concerning first sound from the thermodynamic derivation of the compressibility $\kappa_{T}$ that can be found in Ref.~\onlinecite{AM2011}.  This yields an essentially exact zero-temperature expression for the first sound speed as a function of polarization:
\begin{equation} \label{eq:kappainv}
c_{1}^{2} =  \frac{2 \pi \hbar^{2}}{\overline{n} m^{2}} \biggl[ \left( \frac{m}{m^{*}_{\ua}}\overline{n}_{\ua}^{2}  + \frac{m}{m^{*}_{\da}}\overline{n}_{\da}^{2} \right)  + \tilde{F}_{0}^{\ua \, \ua} \overline{n}_{\ua}^{2} 
			+ 2 \tilde{F}_{0}^{\ua \, \da} \overline{n}_{\ua} \overline{n}_{\da} 
                        	+   \tilde{F}_{0}^{\da \, \da}  \overline{n}_{\da}^{2} \biggr] \,.
\end{equation}
Here $\overline{n}_{\sigma}$ is the areal density in the $\sigma^{\text{th}}$ Fermi sea, $\overline{n} =  \overline{n}_{\ua} + \overline{n}_{\da}$, and $\tilde{F}_{0}^{\sigma \, \sigma'}$ is a dimensionless Landau parameter defined with the bare mass rather than the effective mass. We can rewrite this in a form that is more useful in the present work:
\begin{align} \label{eq:c1sqthermo}
c_{1}^{2}  &= \frac{\left( 1 + \Pee \right)}{4} \Big[ {v_{\text{F}}^{\ua}}^{2} \left( \frac{m^{*}_{\ua}}{m} \right)  [1 + F_{0}^{\ua \ua} ] +  {v_{\text{F}}^{\da}}^{2} \left( \frac{m^{*}_{\da}}{m} \right)  [F_{0}^{\ua \da} ] \Big] \nonumber \\
                 &+ \frac{\left( 1 - \Pee \right)}{4} \Big[  {v_{\text{F}}^{\da}}^{2} \left( \frac{m^{*}_{\da}}{m} \right)  [1 + F_{0}^{\da \da} ] +  {v_{\text{F}}^{\ua}}^{2} \left( \frac{m^{*}_{\ua}}{m} \right)  [F_{0}^{\da \ua} ] \Big] \,.
\end{align}
By inspection of (\ref{eq:c1sqthermo}), in the zero-polarization limit this expression becomes:
\begin{equation}
s_{1}^{2} = \hlf \dfrac{m^{\ast}}{m} (1 + F_{0}^{s}) = \hlf (1 + F_{0}^{s}) (1 + F_{1}^{s}) \,,
\end{equation}
and in the full polarization limit:
\begin{equation} 
s_{1} ^{2} =  \hlf (1 + F_{0}^{\ua \ua}) (1 + F_{1}^{\ua \ua}) \,.
\end{equation}
In the following, these will be referred to as the \textit{thermodynamic} results. 

Thus, both the SBW and AK results in the zero and full polarization limits,  (\ref{eq:fsp0sbw}) and (\ref{eq:fsp0ka}) respectively, are in agreement with the thermodynamic first-sound speeds.   However, as will be seen in Sec.~\ref{sec:he3} there is a slight disagreement at finite polarization $0 < \Pee < 1$ between the thermodynamic sound speed and the SBW/AK results.  

\subsection{\label{ssec:Asym} Asymptotic behavior of zero sound and first sound}

In this section we shall compare the SBW and AK sound speeds and attenuations in the strong and weak-coupling asymptotic limits.  In the strong-coupling limit:  $F_{0} \gg 1$ and $F_{\ell} = 0 \text{ for } \ell \ge 1$.   In the weak-coupling limit, all Landau parameters vanish. Note that for Tables~\ref{tab:zss},~\ref{tab:fss}, and \ref{tab:zsw} $F_{0}^{s}$ is defined as usual with a two spin-state density of states whereas $F_{0}^{\ua \ua}$ is defined with a single spin-state density of states.  

The results for the strong-coupling limit are gathered together in Tables~\ref{tab:zss} and \ref{tab:fss}. These come from analyzing Eqs.~(\ref{eq:zsp0sbw}), (\ref{eq:zsp0ka}), ({\ref{eq:zsp1sbw}), ({\ref{eq:zsp1ka}),     (\ref{eq:fsp0sbw}), (\ref{eq:fsp0ka}), ({\ref{eq:fsp1sbw}), and (\ref{eq:fsp1ka}). The SBW and AK approaches are in exact agreement as far as the sound speeds are concerned. There is a disagreement of $O(1)$ between the two approaches in the coefficients of the attenuation. 
 \begin{table}[h]
\caption{\label{tab:zss} The expressions for the  speed and attenuation of  zero sound in the strong-coupling limit $F_{0} \gg 1$, and $F_{\ell} = 0$ for $\ell \ge 1$ .  }
\begin{ruledtabular}
\begin{tabular}{cccc}
Polarization ($\Pee$) & Approach & Speed squared ($s_{0}^{2}$) & Attenuation ($q_{2}/q_{1}$) \\
\hline
0  & SBW & $\hlf F_{0}^{s}$ & $\frac{1}{2 F_{0}^{s}} \frac{1}{\omega \tau}$ \\
0  & AK   & $\hlf F_{0}^{s}$ & $\frac{5}{2}\frac{1}{F_{0}^{s}} \frac{1}{\omega \tau}$ \\
1 &  SBW & $ \hlf F_{0}^{\ua \ua}$ & $\frac{1}{2 F_{0}^{\ua \ua}} \frac{1}{\omega \tau_{\ua}}$  \\  
1 & AK & $ \hlf F_{0}^{\ua \ua}$ & $\frac{5}{2}\frac{1}{F_{0}^{\ua \ua }} \frac{1}{\omega \tau_{\ua}}$
\end{tabular}
\end{ruledtabular}
\end{table}
 \begin{table}[h]
\caption{\label{tab:fss} The expressions for the  speed and attenuation of  first sound in the strong-coupling limit $F_{0} \gg 1$, and $F_{\ell} = 0$ for $\ell \ge 1$ .   }
\begin{ruledtabular}
\begin{tabular}{cccc}
Polarization ($\Pee$) & Approach & Speed squared ($s_{1}^{2}$) & Attenuation ($q_{2}/q_{1}$) \\
\hline
0  & SBW  & $\hlf F_{0}^{s}$ & $ \frac{\omega \tau}{2 F_{0}^{s}}$  \\
0  & AK  & $\hlf F_{0}^{s}$ & $ \frac{\omega \tau}{4 F_{0}^{s}}$  \\
1 &  SBW & $\hlf F_{0}^{\ua \ua}$ & $ \frac{\omega \tau_{\ua}}{2 F_{0}^{\ua \ua}}$  \\  
1 & AK & $\hlf F_{0}^{\ua \ua}$ & $ \frac{\omega \tau_{\ua}}{4 F_{0}^{\ua \ua}}$
\end{tabular}
\end{ruledtabular}
\end{table}

In the weak-coupling limit, both approaches have the correct ideal Fermi gas  limiting value (at all polarizations).  In Table~\ref{tab:zsw} we point out that the SBW and AK methods approach the limiting values in different ways. The SBW method approaches 1 linearly in the Landau parameter whereas the AK method approaches 1 quadratically. As with the attenuation in the strong-coupling limit, the SBW and AK approaches have slightly different coefficients of the $\omega \tau$ term. 
 \begin{table}[h]
\caption{\label{tab:zsw} The expressions for the  speed and attenuation of  zero sound in the weak-coupling limit $F_{0} \rightarrow 0$, and $F_{\ell} = 0$ for $\ell \ge 1$ . }
\begin{ruledtabular}
\begin{tabular}{cccc}
Polarization ($\Pee$) & Approach & Speed squared ($s_{0}^{2}$) & Attenuation ($q_{2}/q_{1}$) \\
\hline
0  & SBW & $1 + O(F_{0}^{s})$ & $ \frth \frac{1}{\omega \tau}$ \\
0  & AK   & $1 + O((F_{0}^{s})^{2})$ & $\frac{1}{\omega \tau}$ \\
1 &  SBW & $1 + O(F_{0}^{\ua \ua})$ & $\frth  \frac{1}{\omega \tau_{\ua}}$  \\  
1 & AK & $1 + O((F_{0}^{\ua \ua})^{2})$ & $\frac{1}{\omega \tau_{\ua}}$
\end{tabular}
\end{ruledtabular}
\end{table}
 \begin{table}[h]
\caption{\label{tab:fsw} The expressions for the  speed and attenuation of  first sound in the weak-coupling limit $F_{0} \rightarrow 0$, and $F_{\ell} = 0$ for $\ell \ge 1$ .   }
\begin{ruledtabular}
\begin{tabular}{cccc}
Polarization ($\Pee$) & Approach & Speed squared ($s_{1}^{2}$) & Attenuation ($q_{2}/q_{1}$) \\
\hline
0  & SBW  & $\hlf $ & $ \hlf \omega \tau$  \\
0  & AK  & $\hlf $ & $ \frth \omega \tau$  \\
1 &  SBW & $\hlf$ & $ \hlf \omega \tau_{\ua}$  \\  
1 & AK & $\hlf $ & $ \frth \omega \tau_{\ua}$
\end{tabular}
\end{ruledtabular}
\end{table}

\section{ \label{sec:he3} Application to $^{3}$H\lowercase{e} thin films}

In this section we shall compare the results of the SBW and AK approximations as a function of polarization and temperature using Landau parameters that were determined by fitting experimental data for  \he3.    In Refs. \onlinecite{AM2011,LAM_PRB2012} the methods for obtaining these parameters are described in detail.   In very brief  summary, we fit experimental data  for the heat capacity effective mass and the spin susceptibility, obtained for thin \he3 films on graphite substrates.  These fits determine values for the $s$-wave and $p$-wave effective $T$-matrix components at zero polarization.    We then substitute these into expressions for the Landau parameters that are valid to quadratic order in the $T$-matrix components  at arbitrary polarization. The derivations are fairly involved and we refer the reader to the original references for the details. 

In Figs.~\ref{fig:ZsSpeeds}  and  \ref{fig:ZsAtt} we show the zero-sound speed and attenuation at an areal density $\overline{n} = 0.0132$~\AA$^{-2}$ as a function of polarization. We chose  $T = 10$~mK and $\omega = 1500$~MHz  to make sure that the condition for a well-defined zero-sound mode  $\omega \tau \gg 1$ is satisfied at all polarizations. The SBW results are calculated using Eq.~(\ref{eq:csqpm}) directly, while the AK results are obtained using Eq.~(2.32) in Ref.~\onlinecite{LAM_PRB2013}. We note that since we are well into the zero-sound limit, the analytic expressions (\ref{eq:csq}), (\ref{eq:A0sspRe}) and (\ref{eq:A0sspIm}) in the SBW method, and Eqs.~(2.37)-(2.41) in Ref.~\onlinecite{LAM_PRB2013} in the AK method can be used as accurate approximations to the numerical solutions.

Fig.~\ref{fig:ZsSpeeds} shows that the SBW zero-sound speeds are slightly larger than those of AK and the difference is approximately constant for all polarizations ($\sim 5$ m/s at this density). This behavior was noted previously in Ref.~\onlinecite{LAM_PRB2012} (Fig.~9). In Fig.~\ref{fig:ZsAtt} for $0.2 \lesssim \Pee \lesssim 0.9$ the values of the attenuation for both models are fairly close and fairly flat.  Essentially the two models are in reasonable agreement for the zero-sound attenuation. 
\begin{figure}[ht]
\includegraphics[]{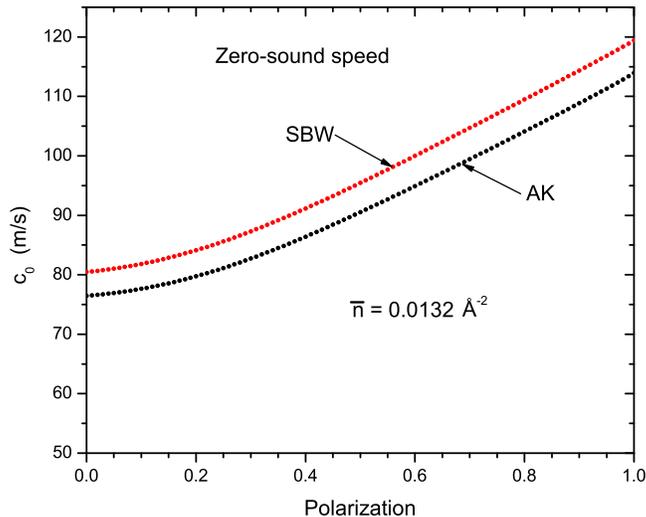}
\caption{(Color online) Comparison of zero-sound speeds as a function of polarization  for the SBW approach  and the AK approach at an areal density $\overline{n} = 0.0132$~\AA$^{-2}$.  The SBW speeds are a fairly constant $\sim 5$~m/s faster than the AK. We note that the two approaches give different values in the limits of zero and full polarization. 
\label{fig:ZsSpeeds}}
\vspace{1truein}
\end{figure}
\begin{figure}[ht]
\includegraphics[]{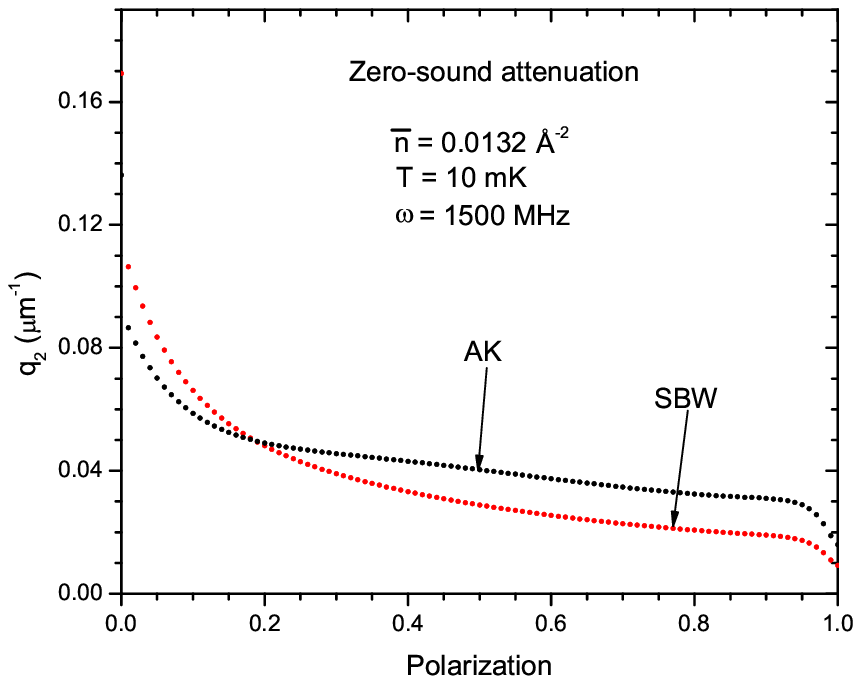}
\caption{(Color online) Comparison of zero-sound attenuation as a function of polarization  for the SBW approach  and the AK approach at an areal density $\overline{n} = 0.0132$~\AA$^{-2}$, frequency $\omega = 1500$~MHz, and temperature $T = 10$~mK.   The results from both methods are very similar. They predict maximum attenuation at zero polarization, and minimum attenuation at full polarization as is to be expected from a simple quasiparticle scattering picture.
\label{fig:ZsAtt}}
\vspace{1truein}
\end{figure}

In Figs.~\ref{fig:FsSpeeds} and \ref{fig:FsAtt} we show the first-sound speed and attenuation at an areal density $\overline{n} = 0.0132$~\AA$^{-2}$ as a function of polarization. We chose  $T = 10$~mK and $\omega = 500$~kHz  to ensure that the condition for a well-defined first-sound mode  $\omega \tau \ll 1$ is satisfied at all polarizations.  Again the SBW results are calculated using Eq.~(\ref{eq:csqpm}), and the AK results are obtained using Eq.~(2.32) in Ref.~\onlinecite{LAM_PRB2013}. In the first-sound limit Eq.~(\ref{eq:c1P}) can be used to evaluate the speed in both models. The attenuation can be evaluated using analytic expressions (\ref{eq:csq}) combined with (\ref{eq:FsApApp}) in the SBW method, or Ref.~\onlinecite{LAM_PRB2013} Eqs.~(2.44)-(2.46) for the AK method. 

We note that in Fig.~\ref{fig:FsSpeeds}, for the first-sound speeds, we also show the \textit{thermodynamic} results from Eq.~(\ref{eq:c1sqthermo}).  As pointed out above: the results for SBW and AK are identical, and at  zero and full polarization SBW/AK agree with the thermodynamic results. However for in-between polarizations $0 < \Pee < 1$ there is a slight disagreement between SBW/AK and Eq.~(\ref{eq:c1sqthermo}).  The \textit{maximum} disagreement, less than 2\%, is certainly less than the uncertainty associated with the values of our Landau parameters. In any case, the possible sources for this disagreement will be discussed in Sec.~\ref{sec:Conclusion}.

The first-sound attenuation calculated by the two models have the same basic behavior: they are minimum at zero polarization and maximum at full polarization.  The factor of two difference in the SBW and AK results can be seen in comparing Eqs.~(\ref{eq:q2SBWP0}) and (\ref{eq:q2AKP0}), and also (\ref{eq:q2SBWP1}) with (\ref{eq:q2AKP1}).  

The qualitative reason for the observed polarization dependence of the  attenuation for both zero sound and first sound can be understood as the consequence of a simple quasiparticle scattering argument.  As shown in Fig.~7 of Ref.~\onlinecite{LAM_PRB2013} the majority spin quasiparticle lifetime increases with increasing polarization. This is mainly due to the decreasing phase space for $s$-wave scattering. Roughly speaking we expect the zero-sound attenuation $\sim 1/ (\omega \tau)$, and thus we expect that zero-sound attenuation should decrease with increasing polarization. Similarly, first-sound attenuation $\sim \omega \tau$, and thus we should expect first-sound attenuation to increase with increasing polarization. Finally, we note that the limiting cases $\Pee \rightarrow 0$ and $\Pee \rightarrow 1$ have technical issues that are discussed in detail in Ref.~\onlinecite{LAM_PRB2013}. We believe that the abrupt behavior seen in these limits in Fig~\ref{fig:ZsAtt} and especially Fig.~\ref{fig:FsAtt} is artificial. 
\begin{figure}[ht]
\includegraphics[]{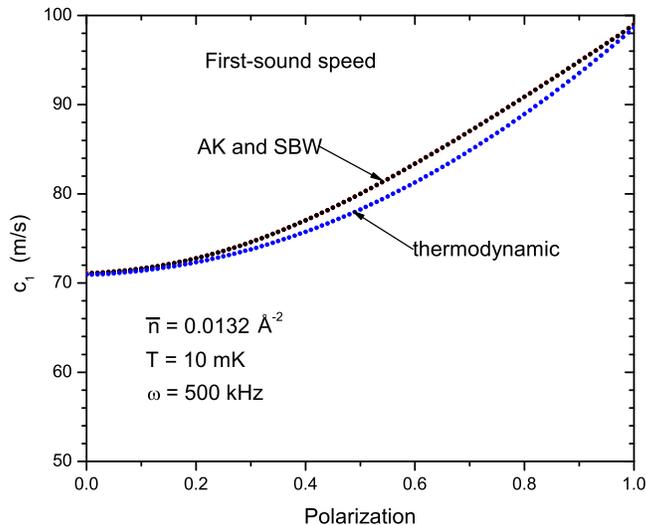}
\caption{(Color online) Comparison of first-sound speeds as a function of polarization  for the SBW approach,  the AK approach, and the exact zero temperature thermodynamic result from Eq.~(\ref{eq:kappainv}).  The areal density is $\overline{n} = 0.0132$~\AA$^{-2}$ and temperature $T = 10$~mK.  The AK and SBW results are in agreement for all polarizations.  The zero-temperature  thermodynamic result Eq.~(\ref{eq:kappainv}) is slightly smaller in magnitude than the AK and SBW.    polarization. 
\label{fig:FsSpeeds}}
\vspace{1truein}
\end{figure}
\begin{figure}[ht]
\includegraphics[]{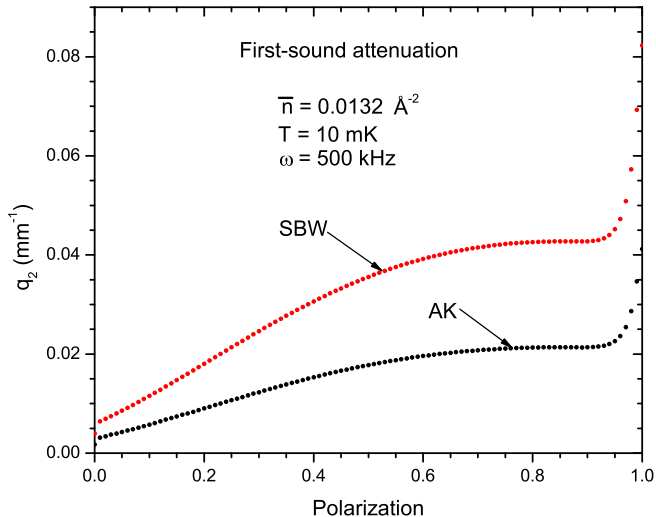}
\caption{(Color online) Comparison of first-sound attenuation as a function of polarization  for the SBW approach  and the AK approach at an areal density $\overline{n} = 0.0132$~\AA$^{-2}$, frequency $\omega = 1500$~MHz, and temperature $T = 10$~mK.   The results from both methods are fairly similar. The abrupt behavior in the regions close to zero polarization and full polarization is probably artificial as discussed in Ref.~\onlinecite{LAM_PRB2013}. 
\label{fig:FsAtt}}
\vspace{1truein}
\end{figure}

In Figs.~\ref{fig:GP00}, \ref{fig:GP05}, \ref{fig:GP10} we show the transition from zero sound to first sound at $\Pee  = 0, 0.5, 1.0$, respectively. For the SBW method the speed and attenuation are obtained by numerically solving Eq.~(\ref{eq:csqpm}), whereas for the AK method they are calculated by numerically solving Eq.~(2.32) in Ref.~\onlinecite{LAM_PRB2013}. We let the temperature increase gradually so that the magnitude of $\omega \tau$ moves smoothly from  $\omega \tau \gg 1$  to $\omega \tau \ll 1$, thus the system transitions continuously from a collisionless zero-sound regime to a hydrodynamic normal sound regime.

In each of these figures the left hand ordinates have their spacing magnified in order to show  clearly the difference in the zero-sound speeds. The small differences between the two methods  decreases as the system moves towards the hydrodynamic regime, and completely vanishes at the first sound limit. The right hand ordinates show the variations of the attenuation over the transition region. The SBW magnitudes are consistently greater but nonetheless still on the same order as the AK ones.   For simplicity we call the \textit{transition temperature} from zero sound to first sound as the temperature that occurs at the peak of the attenuation curve $\omega \tau \approx 1$.  For each polarization  the frequency $\omega$ was adjusted so that the transition temperature is  $\lesssim 10$~mK.  We need to ensure both that the transition temperature is not too low as to not be experimentally accessible, and yet is low enough so that at the highest temperatures needed (in the hydrodynamic regime) the system is still degenerate.  One important feature that these figures show is that for a given frequency, density and polarization, both SBW and AK have their transitions occur at almost the same temperature.  
\begin{figure}[ht]
\includegraphics[]{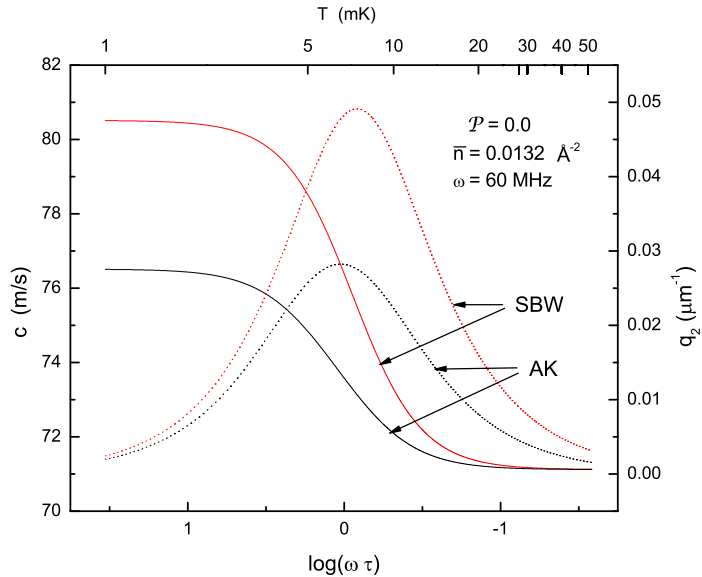}
\caption{(Color online) Sound speed and attenuation as functions of temperature or $\omega \tau$ at zero polarization $\Pee = 0$, density $\overline{n} = 0.0132$~\AA$^{-2}$, and frequency $\omega = 60$~MHz in the region of transition from zero sound to first sound. The SBW and AK approaches yield slightly different results for zero sound but are identical for first sound. The attenuation structures have the same form with different peak heights at the transition. 
\label{fig:GP00}}
\vspace{1truein}
\end{figure}
\begin{figure}[ht]
\includegraphics[]{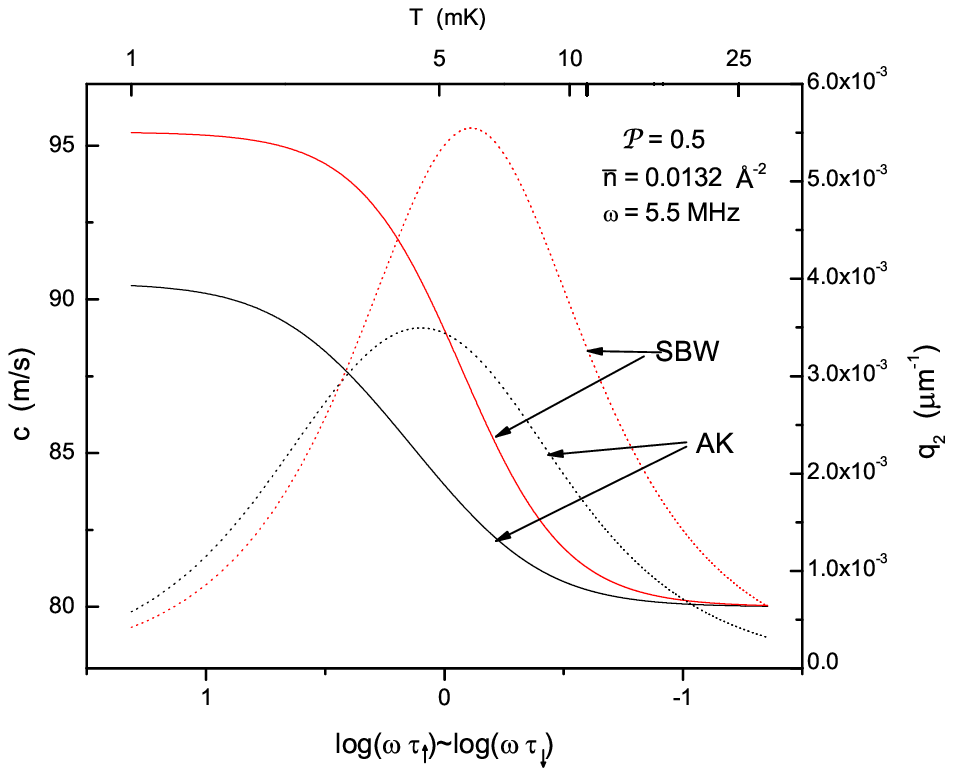}
\caption{(Color online) Sound speed and attenuation as functions of temperature or $\omega \tau$ at polarization $\Pee = 0.5$, density $\overline{n} = 0.0132$~\AA$^{-2}$, and frequency $\omega = 5.5$~MHz in the region of transition from zero sound to first sound. The SBW and AK approaches yield slightly different results for zero sound but are identical for first sound. The attenuation structures have the same form with different peak heights at the transition. 
\label{fig:GP05}}
\vspace{1truein}
\end{figure}\begin{figure}[ht]
\includegraphics[]{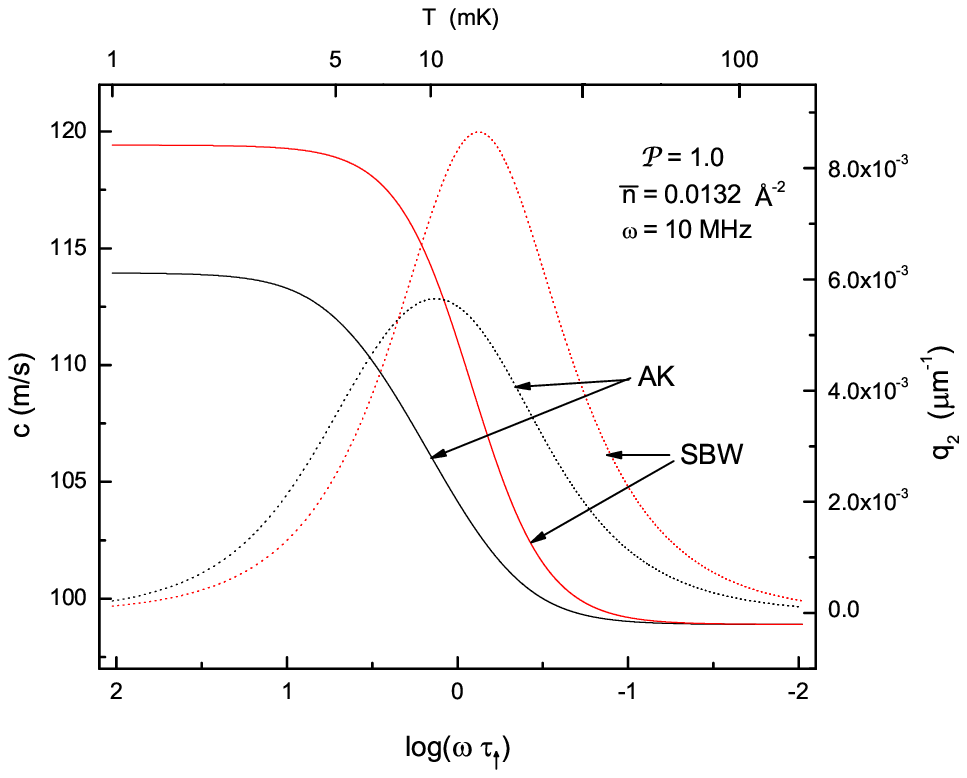}
\caption{(Color online) Sound speed and attenuation as functions of temperature or $\omega \tau$ at full polarization $\Pee = 1.0$, density $\overline{n} = 0.0132$~\AA$^{-2}$, and frequency $\omega = 10$~MHz in the region of transition from zero sound to first sound. The SBW and AK approaches yield slightly different results for zero sound but are identical for first sound. The attenuation structures have the same form with different peak heights at the transition. 
\label{fig:GP10}}
\vspace{1truein}
\end{figure}

\section{Conclusion} \label{sec:Conclusion}
In this manuscript we have explored the solution of Landau's linearized kinetic equation by the method introduced by Sanchez-Castro, Bedell, and Wiegers~\cite{CBW1989} (SBW), and compared it with the solution by using the standard Khalatnikov and Abrikosov~\cite{Khalatnikov1958,*AK1959} (AK) approach.  Both methods rewrite the kinetic equation as an infinite sum of coupled algebraic equations by utilizing a Fourier expansion of all relevant quantities. It is assumed that if both methods retained all terms,  they would yield exact solutions of the kinetic equation. However, for practical solutions both methods need to be truncated, and so one can ask whether the approximate solutions are close to one another, and also whether they are close to exact results. 

There are two fundamental issues that need to be addressed concerning the form and solution of the kinetic equation (\ref{eq:linkep}).  The first concerns the collision integral Eq.~(\ref{eq:CollInt}).  The third term in parentheses was identified as being introduced in order to ensure conservation of momentum.  We note that this is an approximate form for a model with state-dependent lifetimes.  That is, if one requires that total momentum is conserved then the total momentum needs to be computed by summing over the contributions of each constituent. This can easily be done but it leads to an expression in which there are factors of $1 / \tau_{\ua}$ and $1 / \tau_{\da}$ that appear in a numerator and denominator in addition to the overall factor of $1 / \tau_{\ua}$ or $1 / \tau_{\da}$.  This leads to terms that are nonlinear in the collision frequencies. We note that this problem is introduced because of our use of state-dependent lifetimes. If the lifetimes were state independent then the factors of $1 / \tau$ in the numerator and denominator would cancel out. Our choice of (\ref{eq:CollInt}) is essentially equivalent to requiring conservation of momentum for each separate component. In that sense our choice is a sufficient condition for conservation of total momentum. 

The second issue is an important observation made by Troian and Mermin~\cite{TroianMermin1985} concerning the requirement that the collision integral must vanish when the quasiparticle distribution function is in local equilibrium. In order to accomplish this one needs to include the contribution to the quasiparticle energies of the quasiparticle-quasiparticle interaction in the local equilibrium state. Troian and Mermin showed that this inclusion changed the form of the conservation of momentum term by replacing a scalar term by a matrix. They showed that even with a state independent lifetime this change would alter the value of the first sound speed, and bring it into agreement with the exact thermodynamic result. They also pointed out that in the limit of zero polarization and full polarization this change makes no difference since the matrix becomes a $1 \times 1$.  It is possible that we are seeing this effect in the slight difference between the SBW/AK first-sound speeds  and the thermodynamic first-sound speeds for $0 < \Pee < 1$ that we pointed out in Fig.~\ref{fig:FsSpeeds}.   Unfortunately, the Troian-Mermin cure for this problem would involve an analysis similar to the one discussed above concerning conservation of momentum. For \textit{state dependent lifetimes} this would introduce higher order terms in $1/\tau$ than simply linear. We do note that the differences between the SBW/AK and the thermodynamic first-sound speeds are very small: a \textit{maximum} difference of less than 2\% .

In summary we have examined the SBW approach to solving the kinetic equation for complex sound speeds, and compared it with the well-known AK method.   We have calculated analytic expressions for the propagation speeds and attenuation, and also numerical results using Landau parameters appropriate for \he 3 adsorbed on graphite substrates.  For zero-sound speed and attenuation, and first-sound attenuation, the SBW method yields significantly simpler expressions.  Nevertheless, the numerical speeds and attenuation predicted by both methods are very close.  We have noted that in the case of first-sound propagation speeds where SBW and AK yield identical expressions, these results are not in agreement with the thermodynamic results. In the previous section we discussed the possibility that this behavior is due to overlooking of the quasiparticle-quasiparticle interaction contribution to the quasiparticle energy in the distribution function.  In conclusion, we find that the SBW method requires far less algebraic effort than AK to derive analytic expressions for the propagation speeds and attenuation of sound from Landau's kinetic equation, yet nevertheless produces numerical results almost as accurate as the AK method. This is especially true for the zero-sound case at finite polarization $0 < \pol < 1$. Thus, at finite polarization we recommend SBW method as a very attractive alternative to the classic AK.


%

\newpage
%

\end{document}